\documentclass[preprint,12pt]{elsarticle}

\usepackage{amssymb}
\usepackage{amsthm}
\usepackage[mathscr]{eucal}
\journal{Phys. Lett. B}

\begin{document}

\begin{frontmatter}



\title{Dissipationless Hall Current in Dense Quark Matter in a Magnetic Field}
\author{E. J. Ferrer and V. de la Incera}


\address{Dept. of Eng. Sci. and Physics, CUNY-College of Staten Island and CUNY-Graduate Center, 
           New York 10314, USA}

\begin{abstract}
We show the realization of axion electrodynamics within the Dual Chiral Density Wave phase of dense quark matter in the presence of a magnetic field. The system exhibits an anomalous dissipantionless Hall current perpendicular to the magnetic field and an anomalous electric charge density. Connection to topological insulators and 3D optical lattices, as well as possible implications for heavy-ion collisions and neutron stars are outlined.\end{abstract}

\begin{keyword}

Chiral symmetry, Axion QED, Quark-Hole pairing, Cold-Dense QCD, Magnetic DCDW 



\end{keyword}

\end{frontmatter}



\section{Introduction}
In the past few years new macroscopically observable quantum effects that manifest through the interaction of matter with electromagnetic fields in QCD and condensed matter are attracting much attention  \cite{Kharzeev-1501}-\cite{Huang1509}. These effects are connected to the nontrivial topology of these systems and are related to parity and charge-parity symmetry violations. The interaction between the electromagnetic field and matter with nontrivial topology is described by the equations of axion electrodynamics, 

\begin{eqnarray}
\mathbf{\nabla} \cdot \mathbf{E}&=&J_0+\kappa \nabla\theta\cdot \mathbf{B}, \label{axQED1}
 \\
\nabla \times \mathbf{B}-\frac{\partial \mathbf{E}}{\partial t}&=&\mathbf{J}_V-\kappa(\frac{\partial \theta}{dt}\mathbf{B}+\nabla\theta\times \mathbf{E}),  \label{axQED2}
\\
\mathbf{\nabla} \cdot \mathbf{B}&=&0, \quad \nabla \times \mathbf{E}+\frac{\partial \mathbf{B}}{\partial t}=0 \label{axQED3},
\end{eqnarray}  
 proposed by Wilzcek many years ago \cite{axionElect} to describe the effects of adding an axion term $\frac{\kappa}{4}\theta F_{\mu\nu}\widetilde{F}^{\mu\nu}$ to the ordinary Maxwell Lagrangian. In condensed matter, a term of this form has been shown to emerge in: 1) topological insulators (TI) \cite{Qie-PRB78}, where $\theta$ depends on the band structure of the insulator, 2) Dirac semimetals (DM) \cite{Diracsemimet}, a 3D bulk analogue of graphene with non-trivial topological structures, and 3) Weyl semimetals \cite{Weylsemi}, where the angle $\theta$ is related to the energy or momentum separation between the Weyl nodes. 

For quark matter, an electromagnetic axion term can be generated via two separate mechanisms, one at high temperature (T) and the other at high density. At high T, a nontrivial axion field $\theta$ can arise thanks to the sphaleron transitions over the barrier that separates topologically inequivalent vacua \cite{sphaleron}. Even though $\theta$ originally enters coupled to the gluon field, a Fujikawa transformation \cite{FujikawaPRD21_1980} eliminates such a term, but leads to the reappearance of $\theta$ in the QED sector of the theory, where it couples to the electromagnetic field $F_{\mu\nu}$  and its dual \cite{Huang1509}. In the presence of a background magnetic field, the induced $\frac{\kappa}{4}\theta F_{\mu\nu}\widetilde{F}^{\mu\nu}$ term leads to electric charge separation through the well-known Chiral Magnetic Effect (CME) \cite{CME}. The mechanism at high density, takes place in the dual chiral density wave (DCDW) phase of dense quark matter in the presence of a magnetic field. The main purpose of this paper is to  demonstrate the realization of this new mechanism and to show how it leads to the generation of a dissipationless Hall current and an anomalous electric charge.

We highlight from the onset that the role of the magnetic field is quite different at high T than at high density. While at high T the magnetic field just serves, via the CME, as a probe of the nontrivial QCD vacuum topology that produces the axion field $\theta$; in the high-density case, the magnetic field is itself essential to produce the nontrivial topology that manifests, as it will become clear below, through the spectral asymmetry of the quarks in the lowest Landau level (LLL).

\section{Dissipationless Hall current in cold-dense quark matter at $B\neq0$}

 Henceforth, we focus on the cold and dense region of QCD. In this region the CME should be suppressed, an observation consistent with the Beam Energy Scan (BES) results \cite{STAR2}, which show that charge separation starts to diminish already at energies below 60 GeV and disappears completely between 19.6 and 7.7 GeV.  A growing body of works indicates that with increasing density, the chirally broken phase of quark matter is not necessarily replaced by an homogeneous, chirally restored phase, but instead, at least for an intermediate region of densities, the system may favor the formation of inhomogeneous phases. To gain insight on why this occurs notice that with increasing density the homogeneous chiral condensate becomes disfavored due to the high-energy cost of exciting the antiquarks from the Dirac sea to the Fermi surface where the pairs form. At the same time, with higher densities, co-moving quarks and holes at the Fermi surface may pair with minimal energy cost through a mechanism analogous to Overhauser's \cite{Overhauser}, giving rise to a spatially modulated chiral condensate \cite{InhCC}. Spatially modulated chiral condensates in QCD have been discussed in the context of quarkyonic matter \cite{QqM}, where they appear in the form of quarkyonic chiral spirals \cite{QCS} at zero magnetic field, or double quarkyonic chiral spirals  \cite{DQCS} in the presence of a magnetic field. Inhomogeneous chiral condensates have been also studied in the context of NJL models (for a review see \cite{InhCRev}) that share the chiral symmetries of QCD and are then useful to study the chiral phase transition. 
 
From now on, we model cold and dense quark matter in a magnetic field with the help of the NJL-QED Lagrangian density
\begin{eqnarray} \label{L_NJL_QED}
\mathcal{L}=&-&\frac{1}{4}F_{\mu\nu}F^{\mu\nu}+\bar{\psi}[i\gamma^{\mu}(\partial_\mu+iQA_{\mu})+\gamma_0 \mu]\psi 
\nonumber
\\
&+&G[(\bar{\psi}\psi)^2+(\bar{\psi}\mathbf{\tau}\gamma_5\psi)^2],
\end{eqnarray}
with $Q=(e_u,e_d)=(\frac{2}{3}e,-\frac{1}{3}e)$, $\psi^T=(u,d)$; $\mu$ the baryon chemical potential; and G the four-fermion coupling. The electromagnetic potential $A_{\mu}$ is formed by the background $\bar{A}_{\mu}=(0,0,Bx,0)$, that corresponds to a constant and uniform magnetic field $B$ in the z direction, plus the fluctuation field. The presence of the field $B$ favors the formation of the DCDW condensate, 
$\langle\bar{\psi}\psi\rangle+i \langle\bar{\psi}i\tau_3 \gamma_5\psi\rangle= \Delta e^{iqz}$ \cite{KlimenkoPRD82,PLB743} with magnitude $\Delta$ and modulation vector $\mathbf{q}=(0,0,q)$ along the field direction. In this phase, the mean-field Lagrangian is
\begin{eqnarray} \label{DCDW-MF_L}
\mathcal{L}_{MF}&=&\bar{\psi}[i\gamma^{\mu}(\partial_\mu+iQA_{\mu})+\gamma_0\mu]\psi
\nonumber
\\
&-&m\bar{\psi}e^{i\tau_3\gamma_5qz}\psi-\frac{1}{4}F_{\mu\nu}F^{\mu\nu}- \frac{m^2}{4G},
\end{eqnarray}
with $m=-2G\Delta$. The $z$-dependent mass term can be eliminated with the help of the local chiral transformation $\psi \to U_A\psi$, $\bar{\psi} \to \bar{\psi}\bar{U}_A$, with $U_A=e^{-i\tau_3\gamma_5\theta}$, $\bar{U}_A=\gamma_0U^\dag\gamma_0=e^{-i\tau_3\gamma_5 \theta}$, and $\theta(t,\mathbf{x})=\frac{1}{2}q_\mu x^\mu=qz/2$, so that now 
\begin{eqnarray}\label{U_1-MF_L}
\mathcal{L}_{MF}&=&\bar{\psi}[i\gamma^{\mu}(\partial_\mu+iQA_{\mu}-i\tau_3\gamma_5\partial_{\mu}\theta) +\gamma_0\mu-m]\psi 
\nonumber
\\
&-&\frac{1}{4}F_{\mu\nu}F^{\mu\nu}-\frac{m^2}{4G}
\end{eqnarray}
The energy spectrum of the quarks in (\ref {U_1-MF_L}) separates into the LLL  ($l=0$)
\begin{equation}\label{LLLspectrum}
E^{LLL}_k=\epsilon\sqrt{\Delta^2+k_3^2}+q/2,  \quad \epsilon=\pm,
\end{equation}
and the higher ($l>0$) Landau level 
\begin{equation}\label{HighLspectrum}
E^{l>0}_k= \epsilon\sqrt{(\xi\sqrt{\Delta^2+k_3^2}+q/2)^2+2e|B|l}, \quad \epsilon=\pm, \xi=\pm, l=1,2,3,...
\end{equation}
modes. Notice that the LLL spectrum is not symmetric about the zero energy level \cite{KlimenkoPRD82}-\cite{PLB743}.  The asymmetry of the spectrum is characterized by a topological quantity, known as the Atiyah-Patodi-Singer invariant $\eta_B=\sum_k \mathrm{sgn}(E_{k})$ \cite{AS}, a quantity related to the chiral anomaly \cite{PLB743}. This sum is divergent and needs to be properly regularized to ensure that all the energies with equal magnitude and opposite signs cancel out. This implies that only the asymmetric modes contribute to $\eta_B$ and hence the anomalous effects of the system are connected to the LLL. The regularized index $\eta_B=\lim_{s\to0}\sum_k \mathrm{sgn}(E_{k})|E_{k}|^{-s}$ gives rise to an anomalous baryon (quark) number density $ \rho^A_B$. Regularizing the sum with the help of a Mellin transform \cite{PLB743} leads to
 \begin{equation}\label{Baryon_charge}
 \rho^A_B=-N_c\eta_B/2=N_c\sum_{f} \frac{|e_f|}{4\pi^2}\mathbf{B} \cdot \mathbf{\bigtriangledown}(\mathbf{q}\cdot\mathbf{x})=3\frac{|e|}{4\pi^2}qB
 \end{equation}
for the case $q<2m$. The use of a different regularization procedure that allows to extract the anomalous part of the thermodynamic potential, led to the same anomalous quark number density, obtained in this case not from the index $\eta_B$, but as the derivative of the thermodynamic potential with respect to the baryon chemical potential $\mu$ \cite{KlimenkoPRD82}. The extension of this calculation to the isospin asymmetric case was done in \cite{PRD92}.
 When $q>2m$, the quark density acquires an additional, non-topological contribution \cite{PLB743} and becomes 
\begin{equation}
\rho^A_B=-N_c\eta_B/2=-N_c |eB|\left[-q+ \sqrt{q^2-4m^2}\right]/2\pi^2
 \end{equation}
 
Notice that if $B=0$, the quark spectrum is symmetric, so $\eta_B$ vanishes and no anomalous quark number density exists. On the other hand, if $B\ne 0$, but $m=0$, we have $\rho^A_B=0$, consistent with the fact that in this case there is no DCDW condensate, therefore no dependence on the modulation parameter $q$ must remain and no anomalous baryon charge must exist.  These considerations underline that the nontrivial topology of this model results from the interplay of the DCDW ground state and the magnetic field. 

At this point, we should notice that the fermion measure in the path integral is not invariant under the $U_A$ transformation that led to (\ref{U_1-MF_L}), because $\bar{U}_A=U_A \ne U^{-1}_A$ and thus $D\bar{\psi}D\psi \to (\det U_A)^{-2} D\bar{\psi}D\psi$. Hence, it follows that $(\det U_A)^{-2}=e^{-2 \mathrm{Tr} \log U}=e^{-2i\int d^4x \theta(x) \delta^{(4)}(0) \mathrm{tr} \tau_3 \gamma_5}$, with $\mathrm{Tr}$ a functional and matrix trace, and $\mathrm{tr}$ a matrix trace. This expression is ill defined and needs a gauge-invariant regularization. With that aim, we consider a smooth function $f(t)$, such that $f(0)=1$, $f(\infty)=0$, and $tf'(t)=0$ at $t=0$ and $t=\infty$, and regularize the exponent as 
\begin{equation}\label{regularize}
\int d^4x \theta(x) (-2\delta^{(4)}(0) \mathrm {tr} \tau_3 \gamma_5)=-2\lim_{\Lambda\to\infty}\mathrm{Tr} \left[ \theta(x) \tau_3 \gamma_5 f((i D_\mu \gamma^\mu /  \Lambda)^2)\right],
 \end{equation}
where $D_\mu=\partial_\mu+iQA_{\mu}-i\tau_3\gamma_5\frac{q}{2}$ is the corresponding Dirac operator. Following Fujikawa's approach \cite{FujikawaPRD21_1980}, one can show that $(\det U_A)^{-2}_R=e^{i\int d^4x \frac{\kappa}{4}\theta F_{\mu\nu}\widetilde{F}^{\mu\nu}}$. The correct effective Lagrangian is then
\begin{eqnarray} \label{Eff-MFL+axion}
\mathcal{L}_{eff}&=&\bar{\psi}[i\gamma^{\mu}(\partial_\mu+iQA_{\mu}-i\tau_3\gamma_5\partial_{\mu}\theta) +\gamma_0\mu-m]\psi -\frac{m^2}{4G}
\nonumber
\\
&-&\frac{1}{4}F_{\mu\nu}F^{\mu\nu}+\frac{\kappa}{4}\theta F_{\mu\nu}\widetilde{F}^{\mu\nu},
\end{eqnarray}
with $\kappa=\frac{N_c}{2\pi^2}[e_u^2-e_d^2]=\frac{e^2}{2\pi^2}$. The last term in (\ref{Eff-MFL+axion}) is an axion contribution that was overlooked in previous studies of this model. The electromagnetic effective action, found from the partition function after integrating in the fermion fields and expanding in powers of $A_{\mu}$, takes the form
\begin{eqnarray} \label{EA}
\Gamma(A)=&-V\Omega+\int d^4x \left[-\frac{1}{4}F_{\mu\nu}F^{\mu\nu}+\frac{\kappa}{4}\theta F_{\mu\nu}\widetilde{F}^{\mu\nu}\right]
\nonumber
\\
&-\int d^4x A^\mu(x) J_\mu(x)+\cdots,
\end{eqnarray}
with V the four-volume, $\Omega=\Omega(\mu,B)$ the thermodynamic potential \cite{KlimenkoPRD82}, and $J_\mu(x)=(J_0,\mathbf{J})$ the ordinary (nonaxion) electric four-current determined by the sum of the tadpole diagrams of each flavor. 

The linear equations of motion derived from (\ref{EA}) are
\begin{eqnarray}
&\mathbf{\nabla} \cdot \mathbf{E}=J_0+\frac{e^2}{4\pi^2}qB, \label{1}
 \\
&\nabla \times \mathbf{B}-\frac{\partial \mathbf{E}}{\partial t}=\mathbf{J}-\frac{e^2}{4\pi^2} \mathbf{q}\times \mathbf{E},  \label{2}
\\
&\mathbf{\nabla} \cdot \mathbf{B}=0, \quad \nabla \times \mathbf{E}+\frac{\partial \mathbf{B}}{\partial t}=0 \label{3},
\end{eqnarray} 
These are the equations of axion electrodynamics with anomalous electric charge, $J^{anom}_0=\frac{e^2}{4\pi^2}qB$, and current, $\mathbf{J}^{anom}=-\frac{e^2}{4\pi^2} \mathbf{q}\times \mathbf{E}$ densities generated by the axion term in (\ref{EA}). The origin of these anomalous contributions can be  traced back to the asymmetry of the LLL, as can be seen for instance from the connection between the anomalous electric charge and the anomalous quark number density associated with the Atiyah-Patodi-Singer index. Notice that the same expression for the anomalous electric charge density is found if one takes the anomalous quark number density of flavor f, $\rho^A_{Bf}=N_c\frac{|e_f|}{4\pi^2}qB$, multiplies it by the flavor's electric charge $e_f$, and sums in flavor. We highlight that the anomalous current is a dissipationless Hall current, perpendicular to both, the magnetic and the electric field. This could have important consequences for the transport properties of the system. Notice that no CME current is generated here because the axion field $\theta$ is time-independent. 

Before we attempt exploring any implication of the anomalous four-current $J^{anom}_{\mu}$, it is important to be sure that it is not cancelled out by the ordinary four-current $J_\mu(x)$, or more precisely, by its LLL contribution, as the anomalous terms are due to the asymmetric modes of the LLL quarks. To answer this question, we focus on the calculation of the LLL contribution to $J_\mu$, determined by the tadpole diagrams with internal lines of LLL quark propagators.  First, we need to find the LLL propagator of each quark flavor. Keeping in mind that the quarks in the LLL only have one spin projection (parallel/antiparallel to the field for positive/negative charged quarks), and considering $B$ in the positive $z$ direction, we can write the LLL propagator as
\begin{equation}\label{propagator}
G^{\mathrm{sgn}\left(e_f\right)}_{LLL}(p)=D(p,q)\Delta(\mathrm{sgn}\left(e_f\right)),
\end{equation}
with $\Delta(\mathrm{sgn}\left(e_f\right))=(1+\mathrm{sgn}\left(e_f\right) i\gamma^1\gamma^2)/2$ the spin projector, and
\begin{equation}\label{propagator-2}
D(p, q)=\frac{\gamma_\|^\mu \tilde{p}_\mu^-+m}{(\tilde{p}_0^{-})^2-\varepsilon^2}\Delta(+)+\frac{\gamma_\|^\mu \tilde{p}_\mu^++m}{(\tilde{p}_0^{+})^2-\varepsilon^2}\Delta(-),
\end{equation}
Here we used $\tilde{p}_\mu^{\pm}=(p_0-\mu\pm\mathrm{sgn}\left(e_f\right)q, 0, 0, p_3)$, $\gamma_\mu^\|=(\gamma_0,0,0,\gamma_3)$ and $\varepsilon=\sqrt{p_3^2+m^2}$. 

The tadpole diagram for each flavor contributes to the four-current as 
\begin{equation}\label{4-current}
J^\mu_{LLL}(\mathrm{sgn}\left(e_f\right))=(-ie_f)\frac{|e_fB|N_cT}{(2\pi)^3}\sum_{p_4} \int_{-\infty}^{\infty} dp_3 tr \left [i\gamma^\mu G^{\mathrm{sgn}\left(e_f\right)}_{LLL}(p) \right],
\end{equation}
where we have $p_4=\frac{(2n+1)\pi}{\beta}, n=0,1,2,...$, $\beta=1/T$, in the Matsubara sum.

Taking the trace in (\ref{4-current}), we find that the LLL does not contribute to the ordinary electric current density, therefore the Hall current is not eliminated by the ordinary current, an important result that confirms the presence of dissipationless electric transport in the system. 
On the other hand, the LLL contribution to the ordinary electric charge of each quark flavor is  
\begin{equation}\label{4-current-LLL}
J^0_{LLL}(\mathrm{sgn}\left(e_f\right))=\frac{e_f|e_fB|N_cT}{2\pi^2}\sum_{p_4}  \int_{-\infty}^{\infty} dp_3 \frac{\tilde{p}_4-q/2}{(\tilde{p}_4-q/2)^2-\varepsilon^2}, 
\end{equation}
with $\tilde{p}_4=ip_4-\mu$.
Carrying out the Matsubara sum in (\ref{4-current-LLL}), we obtain
\begin{eqnarray}\label{4-current-T}
J^0_{LLL}(\mathrm{sgn}\left(e_f\right))=\frac{-e_f|e_fB|N_c}{2\pi^2} \int_{-\infty}^{\infty} dp_3 & [  n_F(\varepsilon+\mu+q/2)
\nonumber
\\
 &-n_F(\varepsilon-\mu - q/2) ]
\end{eqnarray}
where $n_F(x)=[1+\exp(x)]^{1/2}$ is the Fermi-Dirac distribution.
After integrating, summing in flavor, and taking the zero-T limit, we obtain 
\begin{equation}\label{regular-charge}
J_{LLL}^0=\sum_f J^0_{LLL}(\mathrm{sgn}\left(e_f\right))=\frac{e^2B}{2\pi^2}\sqrt{(\mu-q/2)^2-m^2}\Theta(|\mu-q/2|-m),
\end{equation}
which is the ordinary electric charge density of the LLL quarks in the medium. It is evident that $J^{anom}_0$ is not cancelled out by $J^{LLL}_0$.

\section{Anomalous transport effects in the magnetic DCDW phase}
 
The magnetic DCDW phase exhibits quite interesting properties. The most important is the existence of the dissipationless anomalous Hall current, $\mathbf{J}^{anom}=-\frac{e^2}{4\pi^2} \mathbf{q}\times \mathbf{E}$.  Since it is perpendicular to $\mathbf{E}$ and to the condensate modulation $\mathbf{q}$, which in turn is parallel to $\mathbf{B}$, the Hall current is produced as long as $\mathbf{E}$ and $\mathbf{B}$ are not parallel. A Hall current with these characteristics can appear at the boundary between a topological and a normal insulator \cite{Qie-PRB78} if there is an electric field in the plane of the boundary. In the surface of a TI the anomalous Hall conductance is quantized, but in the magnetic DCDW phase there is no such quantization, and the Hall conductance is $\sigma_H=e^2q/4\pi^2$. Our results are also connected to optical lattices, since a tunable realization of 3D TIs was proposed to exist in 3D optical lattices \cite{PRL105}.

Let us now outline the relevance of these results for neutron stars. Consider a neutron star with a core of DCDW matter threaded by a poloidal magnetic field. Any electric field present in the core, whether due to the anomalous electric charge or not, and as long as it is not parallel to the magnetic field, will lead to dissipationless Hall currents in the plane perpendicular to the magnetic field. Could these currents serve to resolve the issue with the stability \cite{Bstability} of the magnetic field strength in magnetars? How will the electric transport be affected by the anomalous Hall conductance? These and other questions highlight the importance to explore which observable signatures could be identified and then used as telltales of the presence of the DCDW phase in the core. Notice that the condition of electrical neutrality does not need to be satisfied locally for compact hybrid stars \cite{Glendenning}, which could have a core in the magnetic DCDW phase with an anomalous charge contribution and Hall currents circulating inside and at the surface.

The anomalous Hall current could be also produced in future HIC like those planned at the Nuclotron-based Ion Collider Facility (NICA) at Dubna, Russia \cite{PRC85} and at the Facility for Antiproton and Ion Research (FAIR) at Darmstadt, Germany \cite{1607.01487}, which will explore the high density, cold region of the QCD phase map, and where event-by-event off-central collisions will likely generate perpendicular electric and magnetic fields \cite{NICA}.  It will be interesting to carry out a detailed quantitative analysis of how these currents could lead to observable signatures, even after taking into account that there the QGP distributes itself more as an ellipsoid than as an sphere about the center of the collision. The Hall currents will tend to deviate the quarks from the natural outward direction from the collision center and one would expect a different geometry of the particle flow in the DCDW phase compared to other dense phases that have no anomalous electric current. The realization of the DCDW phase in the QGP of future HIC experiments is likely viable because the inhomogeneity of the phase is characterized by a length $\Delta x = \hbar / q \sim 0.6 fm$ for $q \sim \mu =300$ MeV \cite{KlimenkoPRD82}, much smaller than the characteristic scale $L\sim 10 fm$ of the QGP at RHIC, NICA,  and FAIR, while the time scale for this phase will be the same as for the QGP. 

Other interesting effects might emerge by considering the fluctuations $\delta\theta$ of the axion field. If one goes beyond the mean-field approximation, there will be mass and kinetic terms of the axion field fluctuation.  Besides, due to the background magnetic field, the axion fluctuation couples linearly to the electric field via the term $\kappa\delta\theta \mathbf{E}\cdot \mathbf{B}$, so the field equations of the axion fluctuation and the electromagnetic field will be mixed, giving rise to a quasiparticle mode known as the axion polariton mode \cite{axpolariton}. The axion polariton mode is gapped with a gap proportional to the background magnetic field. This implies that electromagnetic waves of certain frequencies will be attenuated by the DCDW matter, since in the DCDW medium they propagate as polaritons. Details of the realization of the axion polariton in the magnetic DCDW medium will be presented in a separate paper. The axion polariton could help to probe the possible realization of the magnetic DCDW in future HIC experiments at high baryon densities, due to its effect in the attenuation of certain light frequencies when light is shined through the collision region.

Notice that both $J^{anom}_0$ and $\mathbf{J}^{anom}$ are proportional to the dynamical parameter $q$, whose value has to be found as a minimum of the thermodynamics potential. This underlines the existence of an interesting connection between ultraviolet (UV) and infrared (IR) phenomena in the DCDW phase in $\mathbf{B}$. Even though quantum anomalies are usually understood as UV phenomena and the UV character of the anomaly in our case is indeed evident from the connection between the axion field $\theta$ and the topological quantity $\eta_H$, we have at the same time that the axion field depends on the modulation parameter q, whose origin is IR because it comes from the quark-hole pairing that is essentially an IR phenomenon. The implication of this connection is that the global topology exhibited by the LLL quarks influences the solution of the inhomogeneous condensate, because the anomaly produces a term in the thermodynamic potential that contributes to the gap equation of q.  A similar UV-IR connection was observed in CME \cite{PPNP75}.

The magnetic DCDW phase exhibits linear magnetoelectricity (ME). This can be seen more easily if we rewrite the first two Maxwell's equations of the medium in terms of the vectors $\mathbf{D}=\mathbf{E}-\kappa\theta \mathbf{B}$ and $\mathbf{H}=\mathbf{B}+\kappa\theta \mathbf{E}$
\begin{equation}
\mathbf{\nabla} \cdot \mathbf{D}=J_0,\quad \nabla \times \mathbf{H}-\frac{\partial \mathbf{D}}{\partial t}=\mathbf{J}_V 
\end{equation} 
Physically this means that a magnetic field induces an electric polarization $\mathbf{P}=-\kappa\theta\mathbf{B}$ and an electric field induces a magnetization $ \mathbf{M}=-\kappa\theta\mathbf{E}$. This is possible because the magnetic DCDW ground state breaks P and T reversal symmetries. The ME here is different from the one found in the magnetic-CFL  phase of color superconductivity \cite{MCFL}, where P was not broken and the effect was a consequence of the anisotropic electric susceptibility \cite{ME-MCFL}, so it was not linear. 

\section{Conclusions}
 
As discussed in this paper, cold and dense QCD in the magnetic DCDW phase exhibits a richness of macroscopically observable quantum topological effects that can be relevant for the HIC physics and neutron stars. It is rather fortunate that we will not have to wait too long to test the realization of these topological phases in the experiment. Future HIC experiments will certainly generate strong magnetic and electric fields in their off-central collisions and will open a much more sensitive window to look into a very challenging region of QCD. For example, the Compressed Baryonic Matter (CBM) at FAIR \cite{1607.01487}  have been designed to run at unprecedented interaction rates to provide high-precision measures of observables in the high baryon density region. That is why it is so timing and relevant to carry out detailed theoretical investigations of all potential observables of the magnetic DCDW phase. Therefore, we hope that our findings will serve to stimulate quantitative studies to identify signatures of the anomalous effects here discussed that will allow to probe the realization of this dense QCD phase both in neutron stars and in HIC experiments. Interestingly, the anomalous effects of the magnetic DCDW phase share many properties with similar phenomena in TIs  and in 3D optical lattices that may even find applications in new devices \cite{axpolariton}. Countertop experiments with optical lattices and TIs  can then help us to gain useful insight of the physics governing the challenging region of strongly coupled QCD. This connection is very promising and will surely continue illuminating and feeding our understanding and intuition in both directions. 

This work was partially supported by the Department of Energy Nuclear Theory grant DE-SC0002179. We thank Igor Shovkovy for interesting discussions and M. Martin-Delgado for calling our attention to the existence of similar physics in 3D optical lattices.

\end{document}